\documentclass{jfm}

\usepackage{graphicx}
\usepackage{natbib}
\usepackage{amsmath}

\usepackage[utf8]{inputenc}

\ifCUPmtlplainloaded \else
  \checkfont{eurm10}
  \iffontfound
    \IfFileExists{upmath.sty}
      {\typeout{^^JFound AMS Euler Roman fonts on the system,
                   using the 'upmath' package.^^J}%
       \usepackage{upmath}}
      {\typeout{^^JFound AMS Euler Roman fonts on the system, but you
                   don't seem to have the}%
       \typeout{'upmath' package installed. jfm.cls can take advantage
                 of these fonts,^^Jif you use 'upmath' package.^^J}%
      }
  \else
  \fi
\fi

\ifCUPmtlplainloaded \else
  \checkfont{msam10}
  \iffontfound
    \IfFileExists{amssymb.sty}
      {\typeout{^^JFound AMS Symbol fonts on the system, using the
                'amssymb' package.^^J}%
       \usepackage{amssymb}%

      }{}
  \fi
\fi

\ifCUPmtlplainloaded \else
  \IfFileExists{amsbsy.sty}
    {\typeout{^^JFound the 'amsbsy' package on the system, using it.^^J}%
     \usepackage{amsbsy}}
    {}
\fi

\newcommand\etal{\mbox{\textit{et al.}}}

\title[Journal of Fluid Mechanics]{Geometry of Valley Growth}

\author[Petroff \etal]{ B\ls y\ns A\ls .\ns P\ls .\ns P\ls E\ls T\ls R\ls O\ls
  F\ls F\ls $^1$,\ns O\ls .\ns D\ls E\ls V\ls A\ls U\ls C\ls H\ls E\ls L\ls
  L\ls E\ls $^1$,\ns D\ls .\ns M\ls .\ns A\ls B\ls R\ls A\ls M\ls S\ls $^{1,2}$,\ns\\
  A\ls .\ns E\ls .\ns L\ls O\ls B\ls K\ls O\ls V\ls S\ls K\ls Y\ls $^1$,\ns
  A\ls .\ns K\ls U\ls D\ls R\ls O\ls L\ls L\ls I\ls $^3$,\ns D\ls .\ns H\ls
  .\ns R\ls O\ls T\ls H\ls M\ls A\ls N\ls $^1$}

\affiliation{ $^1$Department of Earth, Atmospheric and Planetary Sciences,
         Massachusetts Institute of Technology, Cambridge, MA 02139 USA

         $^2$Present address: Department of Engineering Sciences and
         Applied Mathematics, Northwestern University, Evanston, IL
         60208

         $^3$Department of Physics, Clark University, Worcester, MA 01610}

\date{\today}
\volume{000}
\pagerange{\pageref{firstpage}--\pageref{lastpage}}
\doi{???}

\begin{document}
\label{firstpage}
\maketitle

\begin{abstract}
  Although amphitheater-shaped valley heads can be cut by groundwater
  flows emerging from springs, recent geological evidence suggests
  that other processes may also produce similar features, thus
  confounding the interpretations of such valley heads on Earth and Mars.
  To better understand the origin of this topographic form we combine
  field observations, laboratory experiments, analysis of a
  high-resolution topographic map, and mathematical theory to
  quantitatively characterize a class of physical phenomena that
  produce amphitheater-shaped heads.
  The resulting geometric growth equation accurately predicts the
  shape of decimeter-wide channels in laboratory experiments,
  100-meter wide valleys in Florida and Idaho, and kilometer wide
  valleys on Mars.
  We find that whenever the processes shaping a landscape favor the
  growth of sharply protruding features, channels develop
  amphitheater-shaped heads with an aspect ratio of $\pi$.

\end{abstract}

\section{Introduction}

When groundwater emerges from a spring with sufficient
intensity to remove sediment, it carves a valley into the
landscape~\cite[]{dunne1980formation}.
Over time, this ``seepage erosion'' causes the spring to migrate, resulting in an
advancing valley head with a characteristic rounded
form~\cite[]{lamb2006can}.
Proposed examples of such ``seepage channels'' include
centimeter-scale rills on beaches and
levees~\cite[]{higgins1982drainage,schorghofer2004spontaneous},
hundred-meter-scale valleys on
Earth~\cite[]{schumm1995ground,abrams2009growth,russell1902geology,orange1994regular,wentworth1928principles,laity1985sapping},
and kilometer-scale valleys on
Mars~\cite[]{higgins1982drainage,malin1999groundwater,sharp1975channels}.
Although it has long been thought that the presence of an
amphitheater-shaped head is diagnostic of seepage
erosion~\cite[]{higgins1982drainage,russell1902geology,laity1985sapping},
recent work suggests that overland flow can produce similar
features~\cite[]{lamb2006can,lamb2008formation}.
To address this ambiguity, we seek a general characterization of
processes that produce channels indistinguishable in shape from
seepage channels.

We first identify the the interface dynamics
~\cite[]{brower1983geometrical,ben1983dynamics,kessler1985geometrical,shraiman1984singularities,marsili1996stochastic,pelce1988dynamics,pelce2004new}
appropriate for amphitheater-shaped valley heads formed by seepage
erosion.
We then show that the same dynamics apply in a more general setting.
We find that whenever the processes shaping a landscape cause valleys
to grow at a rate proportional to their curvature, they develop
amphitheater-shaped heads with a precise shape.
This result clarifies the controversy surrounding terrestrial and
Martian valleys by showing that many of these features are
quantitatively consistent with a class of dynamics which includes, but
is not limited to, seepage erosion.

%

\section{The Florida network}
\begin{figure}
  \begin{center}
  \includegraphics[width=.9\textwidth]{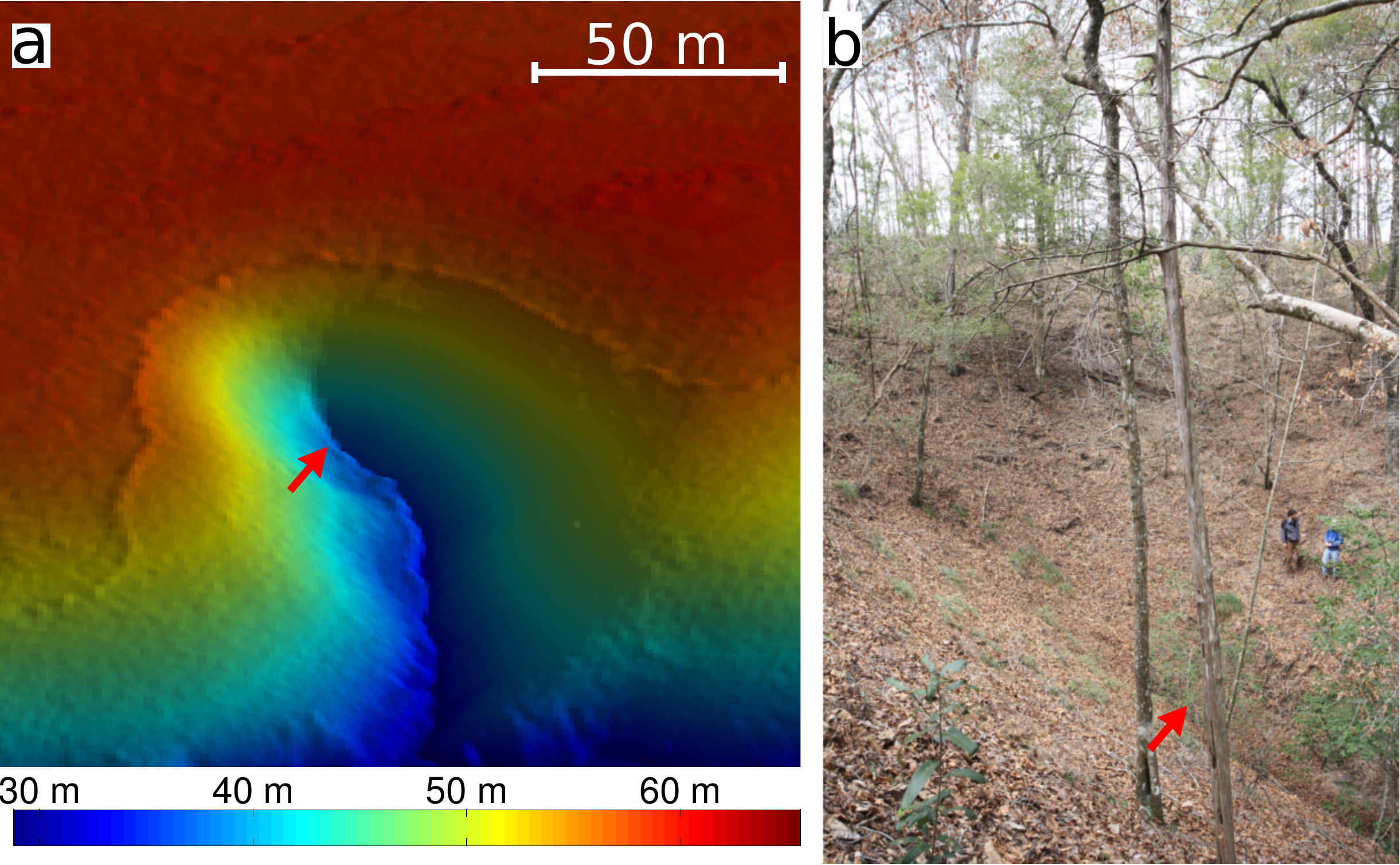}
  \end{center}
   \caption{Examples of seepage valleys from the Florida network. (a)
    Topography obtained from a high-resolution
    map~\cite[]{abrams2009growth} showing the rounded
    ``amphitheater-shaped'' valley head surrounding a spring (red
    arrow). Colors represent elevation above sea level. (b) A seepage
    valley head as viewed from the side wall. The red arrow shows the
    approximate position of the spring. Note people for scale.}
   \label{chaneg}
\end{figure}

To provide a precise context for our analysis, we first focus on a
well-characterized kilometer-scale network of seepage valleys on the
Florida panhandle~\cite[]{schumm1995ground,abrams2009growth}
(figure~\ref{chaneg}).
This network is cut approximately 30 m into homogeneous,
unconsolidated sand~\cite[]{schumm1995ground,abrams2009growth}.
Because the mean rainfall rate $P$ is small compared to the hydraulic
conductivity of the sand, nearly all water enters the channel through
the subsurface~\cite[]{schumm1995ground,abrams2009growth}.
Furthermore, sand grains can be seen moving in streams near the heads,
implying that the water drained by a spring is sufficient to remove
sediment from the head.
Finally, a myriad of amphitheater-shaped valley heads ($n>100$) allows
for predictions to be tested in many different conditions.

We begin by finding the equilibrium shape of the water table in the
Florida valley network. This shape describes how water is distributed
between different heads.
When the groundwater flux has a small vertical component (relative to
the horizontal components), the Dupuit
approximation~\cite[]{bear1979hydraulics} of hydrology relates the
variations in the height $h$ of the water table above an impermeable
layer~\cite[]{schumm1995ground,abrams2009growth} to the mean rainfall
rate $P$ and the hydraulic conductivity $K$ through the Poisson
equation
\begin{equation}
  \frac{K}{2}\nabla^{2}h^{2}+P=0.
  \label{poisson}
\end{equation}
%
%
To simplify our analysis, we define two rescaled quantities: the
Poisson elevation $\phi=(K/2P)^{1/2}h$ and the Poisson flux
$q_{p}=\|\nabla\phi^{2}\|$.
The Poisson elevation is determined entirely from the shape of the
network and, consequently, can be measured from a map without
knowledge of $P$ or $K$.
Physically, $q_{p}$ is the area that is drained by a small piece of
the network per unit arc length.
It is therefore a local version of the inverse drainage density (i.e.,
the basin area divided by total channel length).
%
%
By construction, the groundwater flux $q=P q_{p}$.
This measure of the flux differs from the geometric drainage
area~\cite[]{abrams2009growth} in that it is found from a solution of
the Poisson equation, rather than approximated as the nearest
contributing area.

\begin{figure}
  \begin{center}
  \includegraphics[width=.85\textwidth]{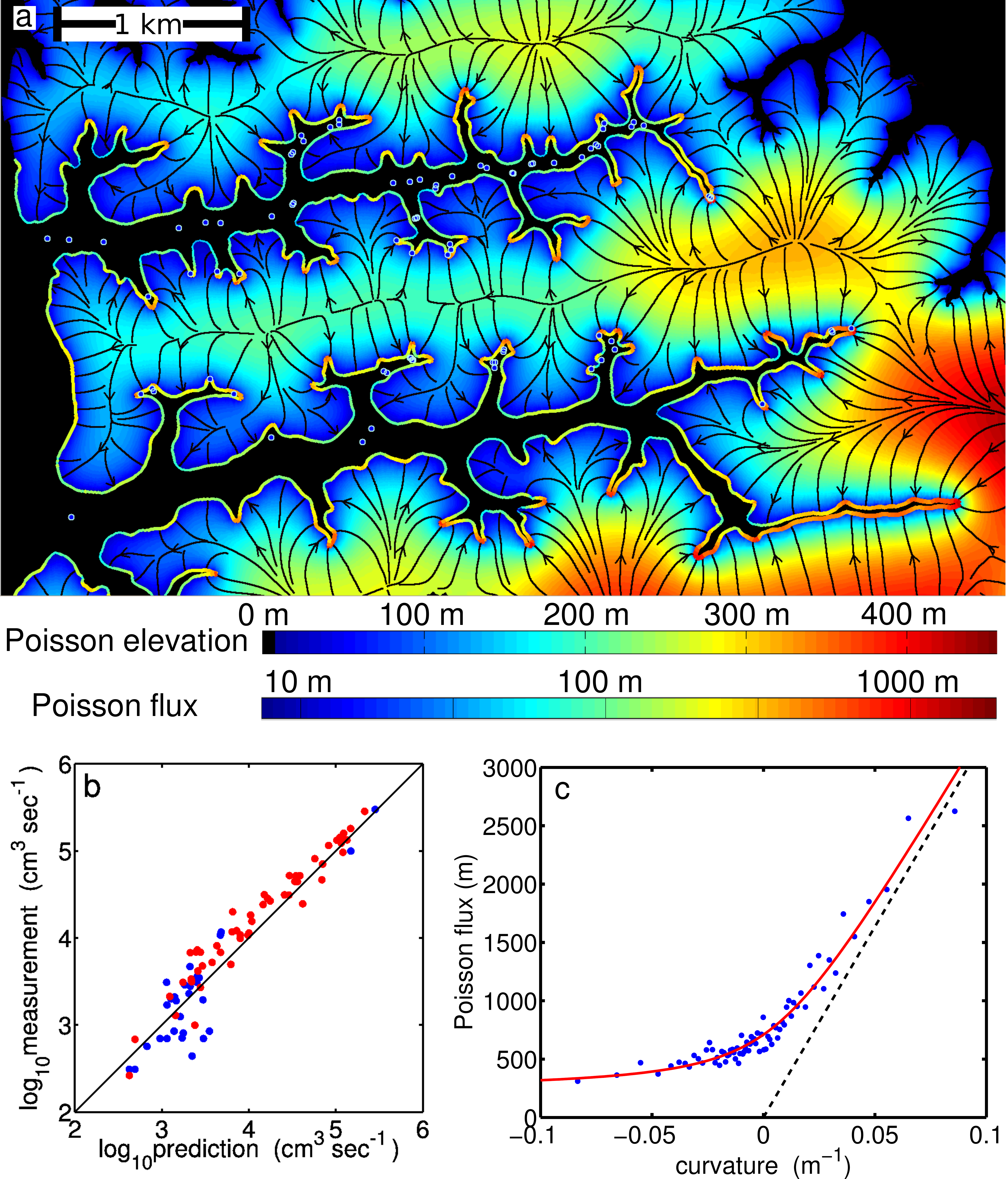}
  \end{center}
  \caption{The water table and associated groundwater flux in the
    Florida network. (\textbf{a}) The magnitude of the Poisson flux
    (color intensity on boundary) is the size of the area draining
    into a section of the network per unit length. It is found by
    solving equation~(\ref{poisson}) around the channels as
    approximated with an elevation contour. Flow lines are in
    black. The water discharge was measured at blue circles. The
    Poisson elevation and Poisson flux are proportional to the water
    table height and groundwater flux, respectively. (\textbf{b})
    Comparison of the predicted discharge to measurements at 30 points
    in network taken in January of 2009 (blue points) and 52 points in
    April of 2009 (red points). The black line indicates
    equality. This comparison is direct and requires no adjustable
    parameters. (\textbf{c}) We observe a hyperbolic relationship
    between the curvature of the valley walls and the predicted flux
    (red curve). In regions of high curvature (i.e. valley heads) the
    flux is proportional to curvature (dashed line).}
  \label{fig1}
\end{figure}

Fig.~\ref{fig1}a shows the solution of equation~(\ref{poisson}) around
the valley network (supplementary material). Because the variations in
the elevation at which the water table emerges are small ($\sim10$~m)
relative to the scale of the network ($\sim1000$~m), we approximate
the network boundary with an elevation contour extracted from a high
resolution topographic map~\cite[]{abrams2009growth} on which $\phi$ is
constant (see supplementary material).
For a specified precipitation rate, this result predicts the flux $q$
of water into each piece of the network.

To test this model of water flow, we compared the solution of
equation~(\ref{poisson}) to measurements at 82 points in the network.
Given a reported mean rainfall rate of
$P=5\times10^{-8}$~m~sec$^{-1}$~\cite[]{abrams2009growth}, we find good
agreement between observation and theory (Fig.~\ref{fig1}b),
indicating that equation~(\ref{poisson}) accurately describes the
competition for groundwater.
Additionally, we find that the water table elevation $h$ is consistent
with a ground penetrating radar survey~\cite[]{abrams2009growth} of the
area (see supplementary material).
To understand how the distribution of groundwater through the network
produces channels with amphitheater-shaped heads, we proceed to
relate the flux of water into a valley head to the geometry of the
head.


\section{Relation of flux and geometry}

For an arbitrary network, there is no simple relationship between the
flux of water into part of the network and its local shape.
As each tip competes with every other part of the network, one can
only find the local flux by solving equation~(\ref{poisson}).
However, as first identified by Dunne~\cite[]{dunne1980formation},
valleys cut by seepage grow when sections of the valley which
protrude outwards (high positive curvature) draw large fluxes while
indented sections (negative curvature) of the network are shielded by the
network.
Motivated by this insight, we seek the relationship between the flux
into a piece of a valley network and its planform curvature.
Fig.~\ref{fig1}c shows that this relationship is consistent with a
hyperbolic dependence of the Poisson flux (and hence the water flux)
on the curvature.
%
%
Consequently, at tips, where the curvature is high, this relationship
can be approximated by the asymptote. Thus,
\begin{equation}
  q_{p}\simeq\Omega\kappa,
  \label{hyperbola}
\end{equation}
where $\Omega$ is a proportionality constant related to the area
drained by a single head.
Thus we find a local relationship between the processes shaping a
seepage valley, represented by the flux $q_{p}$, and the local
geometry, represented by the curvature $\kappa$.
We note that this relation may be further justified by a scaling
argument (supplementary material), but here we merely employ it as
an empirical observation.

%

\section{Geometric growth law}
In what follows we first ask how the flux-curvature
relation~(\ref{hyperbola}) is manifested in the shape of a single
valley head.
To do so we first find the shape of a valley head that is consistent
with the observed proportionality between groundwater flux and
curvature.
This derivation relies on three steps.
First, equation~(\ref{hyperbola}) is converted, with an additional
assumption, into a relationship between the rate at which a valley
grows outward and its planform curvature.
Next, we restrict our attention to valley heads that grow forward
without changing shape.
This condition imposes a geometric relationship between growth and
orientation.
Combining these, we find a relationship between curvature and
orientation that uniquely specifies the shape of a valley growing
forward due to groundwater flow.
Finally, we find that our theoretical prediction is consistent both
with valleys cut by seepage and systems in which seepage is doubtful.
This result leads us to conclude that seepage valleys belong to a
class of systems defined by a specific relationship between growth and
curvature which includes seepage erosion as a particular case.

%
Following past work~\cite[]{howard1988groundwater,abrams2009growth}, we
assume that the amount of sediment removed from the head is
proportional to the flux of groundwater; and thus, by
equation~\ref{hyperbola}, $\Omega \kappa$.
From equation~(\ref{hyperbola}), the speed $c$ at which a valley grows
outward is therefore proportional to the planform valley curvature
$\kappa$. Setting $H$ equal to the difference in elevation between the
spring and the topography it is incising, the sediment flux is
\begin{equation}
  Hc=\alpha \Omega \kappa,
\label{kappac}
\end{equation}
where $\alpha$ is a proportionality constant with units of velocity.
Equation~(\ref{kappac}) states that the more sharply a valley wall is
curved into the drainage basin, the faster it will grow. The growth of
the channel head is therefore ``curvature-driven''
~\cite[]{brower1983geometrical}.

This derivation of equation~\ref{kappac} marks a shift of focus from the
mechanics that shape a seepage valley to the dynamics by which it
evolves.
Although the specific processes of groundwater flow and sediment
transport have not been addressed explicitly, this generalization has
two advantages.
First, equation~\ref{kappac} is purely geometric and can be solved to provide a
quantitative prediction for the shape of a valley head.
Equally importantly, the generality of these dynamics suggests that
the class of processes they describe may extend beyond seepage valleys
and thus provide a quantitative prediction for the evolution of a
wider class of channelization phenomena.

\section{Shape of a valley head}

\begin{figure}
 \begin{center}
  \includegraphics[height=9cm]{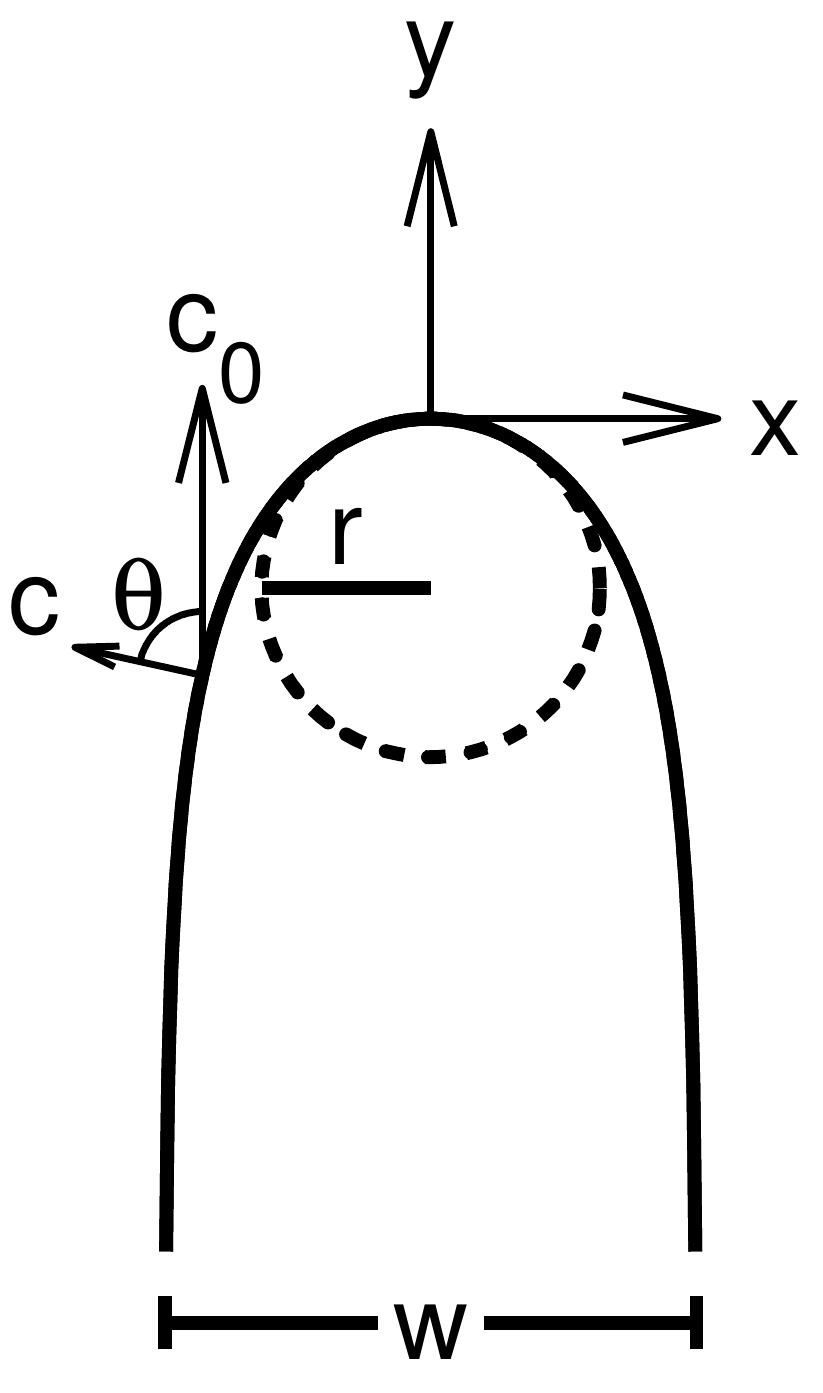}
 \end{center}
  \caption{A balance between curvature-driven growth and translational
    growth sets the shape (eq.~\ref{logcos}) of an amphitheater-shaped
    valley head (solid black curve). When a curve evolves due to
    curvature-driven growth, the normal velocity $c$ is inversely
    proportional to the radius of the best fitting circle at that
    point. When a curve translates forward, there is a geometric
    relationship between the speed at which a point translates $c_{0}$
    and the speed at which it grows in the normal direction $c$. This
    balance produces channels with a well defined width $w$ and an
    aspect ratio of $w/r=\pi$.}
  \label{logcos_eg}
\end{figure}

We restrict our attention to steady-state valley growth.
When the channel grows forward at a speed $c_{0}$ without changing
shape, the outward growth balances the growth forward.
If $\theta$ is the angle between the normal vector and the direction
the channel is growing (Fig.~\ref{logcos_eg}), then
$c=c_{0}\cos\theta$.
Substituting this condition for translational growth into equation~\ref{kappac}
relates the orientation of a point on the channel to the curvature at
that point:
\begin{equation}
  \cos\theta=\frac{\alpha \Omega}{c_{0}H}\kappa(\theta),
  \label{kappa_theta1}
\end{equation}
where $\kappa(\theta)$ denotes the dependence of curvature on
orientation. 
Solving this equation (see appendix) for the shape of the curve with
this property gives~\cite[]{brower1983geometrical}
\begin{equation}
  y(x)=\frac{w}{\pi}\log\cos\left(\pi\frac{x}{w}\right),
  \label{logcos}
\end{equation}
where $w=\pi \alpha \Omega/(c_{0}H)$ is the valley width and
$\theta=\pi x/w$. 
The planform shape $y(x)$ is shown in Fig.~\ref{logcos_eg}.
A notable feature of this solution is that all geometric aspects of
the channel head are set by the absolute scale of the valley (i.e. the
valley width).
In particular, it follows from equations~(\ref{kappa_theta1}) and
(\ref{logcos}) that all seepage channels have a
characteristic aspect ratio
\begin{equation}
  \frac{w}{r}=\pi,
  \label{aspectratio}
\end{equation}
where $r$ is the radius of curvature of the tip
(Fig.~\ref{logcos_eg}).
By contrast, a semi-circular valley head, in which $w=2r$, has an
aspect ratio of 2.

\section{Comparison to observation}

To test these predictions, we first compare the shape of elevation
contours extracted from 17 isolated, growing tips in the Florida
network to equations~(\ref{logcos}) and (\ref{aspectratio}).
As these valley heads vary in size, a sensible comparison of their
shapes requires rescaling each channel to the same size; we therefore
non-dimensionalize each curve by its typical radius $w/2$.
To remove any ambiguity in the position where the width is measured,
$w$ is treated as a parameter and is fit from the shape of each valley
head.
Fig.~\ref{fig4}a compares all 17 rescaled channels heads to
equation~(\ref{logcos}).
Although each individual valley head may deviate from the
idealization, the average shape of all valley heads fits the model
well.

This correspondence between theory and observation is further
demonstrated by comparing the average curvature at a point to its
orientation.
We construct the average shape of the valley head by averaging the
rescaled contours along the arc length.
Rewriting equation~(\ref{kappa_theta1}) in terms of the width, we
obtain
\begin{equation}
w \kappa=\pi\cos\theta.
\label{kappa_theta}
\end{equation}
Plotting $w\kappa$ as a function of $\cos\theta$, we indeed observe
this proportionality (Fig.~\ref{fig4}b). Moreover the measured slope
$p=3.07\pm0.17$ is consistent with the predicted prefactor $p=\pi$.
The proportionality relationship holds most clearly at high
curvatures, where the approximation that flux scales with curvature is
most accurate.
Notably, were amphitheater-shaped valley heads semi-circular, then
Fig.~\ref{fig4}b would show the horizontal line $w\kappa(\theta)=2$.
If valley heads were sections of an ellipse with an aspect ratio of
$\pi$, the data in Fig.~\ref{fig4}b would follow the curve $w\kappa(\theta)
=(4+(\pi^{2}-4)\cos^{2} \theta)^{3/2}/\pi^{2}$.
Viewing the semi-circle and ellipse as geometric null hypotheses, we
conclude from visual inspection of Fig.~\ref{fig4}b that we can
confidently reject them in favor of equation~(\ref{logcos}).

Seepage channels can also be grown in the laboratory by forcing water
through a sand
pile~\cite[]{schorghofer2004spontaneous,howard1988groundwater,lobkovsky2007erosive}.
Because these channels grow on the time scale of minutes to hours, one
can directly observe the development of the channels.
Fig.~\ref{fig4}c compares equation~(\ref{logcos}) to elevation
contours extracted from a previous
experiment~\cite[]{lobkovsky2007erosive} while the channel is growing.
Once the contours are rescaled and averaged, the curvature again is
proportional to $\cos\theta$ (Fig.~\ref{fig4}d). The measured
proportionality constant $p=3.07\pm0.13$, consistent with $p=\pi$.

%
%
%
%
%

\section{Generalizations}

\begin{figure}
  \begin{center}
  \includegraphics[width=1\textwidth]{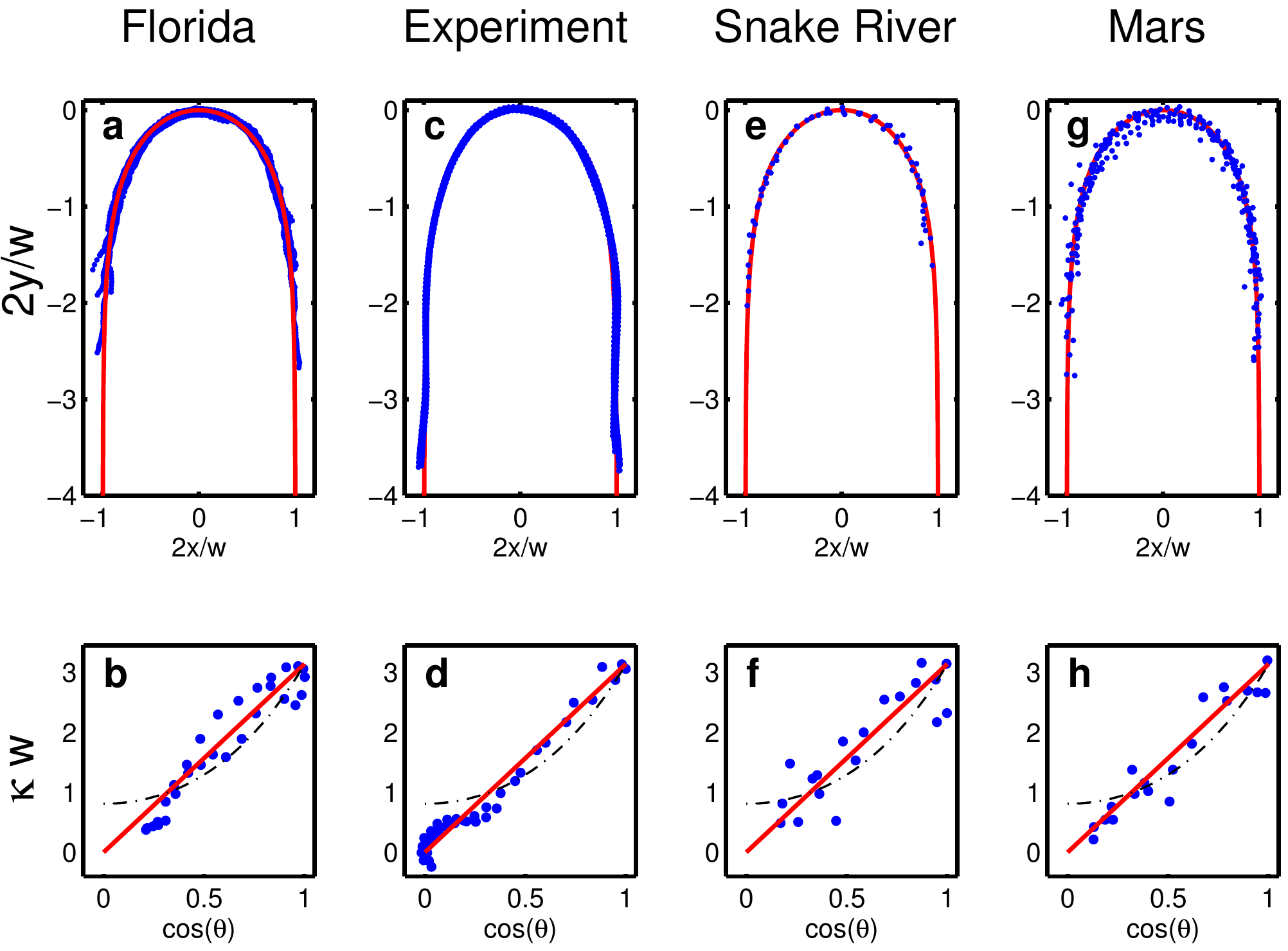}
  \end{center}
  \caption{The shape of valley heads in the field, experiments, on
    Earth, and on Mars are consistent with curvature-driven growth.
    (\textbf{a}) The shape of a channel produced by curvature-driven
    growth (red line) compared to the relative positions of points
    (blue dots) on the edge of valleys from the Florida network (17
    elevation contours).
    (\textbf{b}) Comparison of the curvature at a point to the
    orientation (blue dots) of valleys from the Florida network.  The
    red line is the linear relationship given in
    equation~(\ref{kappa_theta}). The black dashed line corresponds to
    an ellipse with aspect ratio $\pi$. A semi-circular head would
    predict the horizontal line $\kappa w=2$.
   (\textbf{c}--\textbf{d}) The analogous plots for the experiments
    (25 elevation contours extracted at 3 minute intervals).
   (\textbf{e}--\textbf{f}) The analogous plots for three valleys near
    Box Canyon and Malad Gorge.
   (\textbf{g}--\textbf{h}) The analogous plots for 10 Martian
    ravines.}
  \label{fig4}
\end{figure}

The strong correspondence between equation~(\ref{kappa_theta}) and the
observed shapes of valley heads suggests that amphitheater-shaped
heads take their form from curvature-driven growth.
Because curvature-driven growth is a simple geometric growth
model, it likely characterizes a class of physical
processes~\cite[]{brower1983geometrical}.
For example, when a low-viscosity fluid is pushed slowly into a
viscous fluid in two-dimensions, the diffusing pressure field deforms
the intruding fluid into an elongated form as described by the
Saffman-Taylor instability~\cite[]{saffman1958penetration}.
When stabilized by surface tension, the shape of the resulting
``viscous finger'' is exactly that given in
equation~(\ref{logcos})~\cite[]{bensimon1986viscous,combescot1986shape}.
This morphology has also been related to the shape of
dendrites~\cite[]{mullins1963morphological,kessler1986dendritic} and is a
steady state solution to the deterministic Kardar--Parasi--Zhang
equation~\cite[]{kardar1986dynamic}.

This generality leads us to conjecture that when the growth of a
valley head responds linearly to a diffusive flux, its dynamics at
equilibrium reduce to curvature-driven growth.
Geophysically relevant processes in which the growth may be dominated
by a (possibly non-linear) diffusive flux include the conduction of
heat, topographic diffusion~\cite[]{culling60}, the shallow water
equations~\cite[]{chanson1999hydraulics}, and elastic
deformation~\cite[]{landau1995theory}.
Thus, assuming appropriate boundary conditions exist, processes such
as seasonal thawing, the relaxation of topography, overland flow, and
frost heave may produce valleys indistinguishable in planform shape
from seepage channels.


To confirm the wide applicability of he geometric growth model, we
proceed to compare equations~(\ref{logcos}) and (\ref{kappa_theta}) to
enigmatic valleys on Earth and Mars.
%
%
The origins of amphitheater-shaped heads from the Snake River in
Idaho~\cite[]{russell1902geology,lamb2008formation} and the Martian
valleys of Valles Marineris have been the subject of much
debate~\cite[]{higgins1982drainage,malin1999groundwater,sharp1975channels,lamb2006can}.
Fig.~\ref{fig4}(e-h) shows that the shape of valley heads in both of
these systems is consistent with equations~(\ref{logcos}) and
(\ref{kappa_theta}).
Averaging the rescaled valleys and comparing the dimensionless
curvature to the orientation, we find $p=2.92\pm0.24$ and
$p=3.02\pm0.21$ for the Snake River and Martian features respectively.
Both estimates are consistent with $p=\pi$.

\section{Conclusion}
That these valleys are consistent with the predictions of
curvature-driven growth does not, however, necessarily imply that
their growth was seepage-driven.
We favor instead a more conservative conclusion: diffusive transport is
ubiquitous and therefore so too is the $\log\cos\theta$ form.

%


Our results clarify the debate about the origin of amphitheater-shaped
valley heads by placing them within a class of dynamical phenomena
characterized by growth proportional to curvature.  From this
qualitative distinction we obtain a quantitative prediction: the
valley head has a precisely defined shape with an aspect ratio of
$\pi$.  Regardless of the specific mechanical processes that cause a
particular valley head to grow, all valley heads that fall within this
dynamical class will look alike. 

\bigskip

We would like to thank The Nature Conservancy for access to the
Apala{\-}chicola Bluffs and Ravines Preserve, and K. Flournoy,
B. Kreiter, S. Herrington and D. Printiss for guidance on the
Preserve. We thank B. Smith for her experimental work. It is also our
pleasure to thank M. Berhanu. This work was supported by Department of
Energy Grant FG02-99ER15004. O. D. was additionally supported by the
French Academy of Sciences.

\bibliography{chan_bib}
\bibliographystyle{jfm}

\end{document}


\begin{center} {\Large {\bf Supplementary Material} \vskip 1em {\em
      for} \vskip 1em {\bf Geometry of Valley Growth}}
\end{center}

{\parindent 0pt
\begin{center}
  {Alexander P. Petroff$^1$, Olivier Devauchelle$^1$, Daniel M. Abrams$^{1,2}$,\\
    Alexander E. Lobkovsky$^1$, Arshad Kudrolli$^3$, Daniel H. Rothman$^1$}
\end{center}
  \vskip 2em}

{\parindent 0pt
  \vskip 2em

  {\sl   $^1$Department of Earth, Atmospheric and Planetary Sciences,
         Massachusetts Institute of Technology, Cambridge, MA 02139 USA

         $^2$Present address: Department of Engineering Sciences and
         Applied Mathematics, Northwestern University, Evanston, IL
         60208

         $^3$Department of Physics, Clark University, Worcester, MA 01610
       }}

\tableofcontents

\newpage

%
%
\section{Computation of the water table}

In order to find the distribution of groundwater flux into the
network, we solved for the shape of the water table around the
channels.
From the main text, the Poisson elevation $\phi$ of the water table is
a solution to the equation:
\begin{equation}
  \nabla^{2}\phi^{2}+1=0
\label{poisson_}
\end{equation}
with absorbing and zero flux boundary conditions. Thus $\phi$ is
independent of the hydraulic conductivity $K$.

The ground water flux at a point is related to the shape of the
watertable through the equation
  \begin{equation}
    q=\frac{K}{2}\|\nabla h^{2}\|
  \end{equation}
  from which
  \begin{equation}
    q=P\|\nabla \frac{K}{2 P}h^{2}\|
  \end{equation}
  Thus, from the definition of $\phi$
   \begin{equation}
    q=P\|\nabla\phi^{2}\|
  \end{equation}
  Because $\phi$ is only a function of the network geometry, $q$ is
  independent of $K$. This result also follows from conservation of
  mass. The total discharge from the network must be equal to the
  total rain that falls into the network, regardless of
  conductivity. $K$ sets the slope of watertable at the boundary
  required to maintain this flux.

\section{Selection of the boundary}

\begin{figure}[h]
\centering
\includegraphics[scale=1]{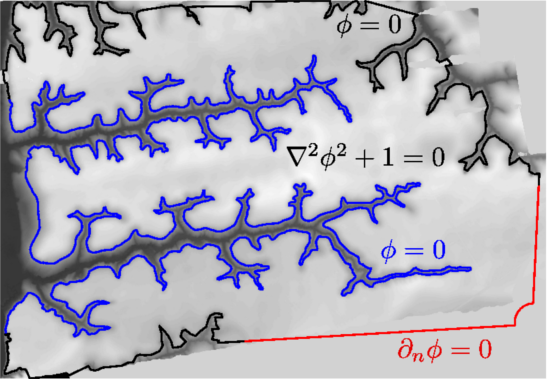}
\caption{To most closely approximate the shape of the network we use
  an elevation contour of the topography. Approximating the channels
  as nearly flat, we required that the water table intersect the
  channels at a constant height, which we chose as zero. This boundary
  is drawn in blue and black. Additionally, a drainage divide (red
  line) was included in the south east. Because our LIDAR
  map~\cite{abrams2009growth} only shows two full valley networks, we
  only analyze the data from this portion of the boundary (blue
  line). The boundary is linearly interpolated between points spaced
  at 20 m intervals on the blue boundary and points spaced by an
  average of 50 m on the red and black boundaries.}
\label{boundary}
\end{figure}

We solve the equation around a boundary chosen to follow the position
of springs and streams. To identify such a boundary, we first remove
the mean slope ($0.0025$) of the topography. We then chose the 45 m
elevation contour of the resulting topography as the boundary
(Figure~\ref{boundary}) obtained from a high resolution LIDAR map of
the network~\cite{abrams2009growth}. This elevation was chosen as the
approximate elevation of many springs. When the contour exits the area
where the LIDAR map was available, we replace the missing section of
the channel with an absorbing boundary condition. Because this
approximation results in uncertainties in the flux near the missing
boundary, we only analyze the water flux into a well contained section
of the network (blue boundary in Figure~\ref{boundary}). Finally, we
include a zero-flux boundary condition in the south east in the
approximate location of a drainage divide. We solve
equation~(\ref{poisson_}) with these boundary conditions using a
finite-element method~\cite{hecht2005freefem++}.

\section{Comparison of the shape of the water table to the Poisson
  elevation}

\begin{figure}
  \centering
\begin{tabular}{cc}
\includegraphics[scale=.4]{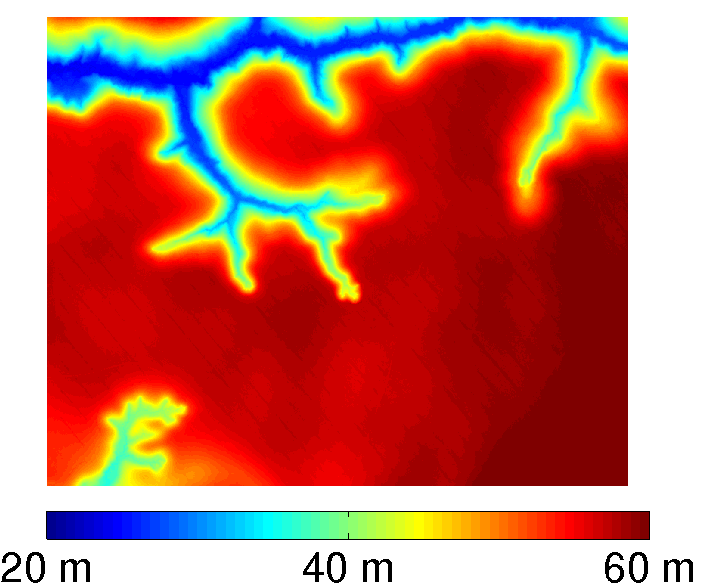}&
\includegraphics[scale=.4]{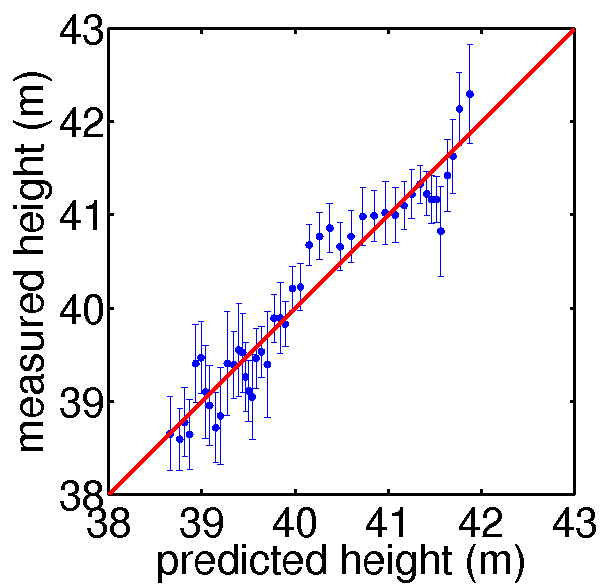}\\
(a)&(d)\\        
\includegraphics[scale=.4]{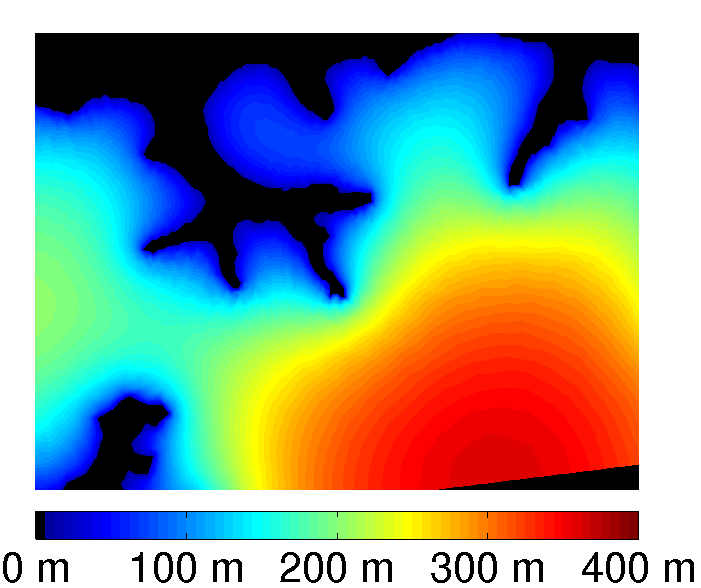}&
\includegraphics[scale=.4]{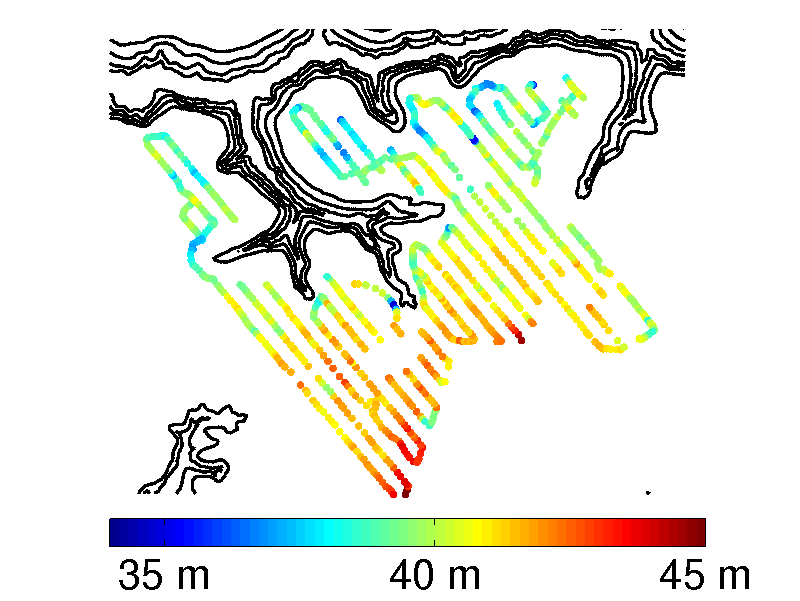}\\
(b)&(c)\\
\end{tabular}
\caption{Comparison of the Poisson elevation to field observation. (a)
  The available ground penetrating radar survey was conducted on a
  portion of the southern valley network. The topographic map of the
  channels near the survey is 1400 m across. (b) We solved
  equation~(\ref{poisson_}) around the valley for the Poisson
  elevation. (c) The ground penetrating radar
  survey~\cite{abrams2009growth} provided the elevation of the water
  table above sea level at 1144 points around the network. The valley
  walls are represented by the elevation contours for 30 m to 45 m at
  5 m intervals. (d) The measured height is consistent with
  theory. The red line indicates perfect agreement.}
\label{GPRfig}
\end{figure}

Here we show that the solution of equation~(\ref{poisson_}) is consistent with
field observations.
We compare $\phi$ (Figure~\ref{GPRfig}b) to a previously
reported~\cite{abrams2009growth} ground penetrating radar (GPR)
survey of the channels (Figure~\ref{GPRfig}c).
%

%
As all heights are measured relative to the impermeable layer, we
define $h_{0}$ to be the reference elevation and shift $h$
accordingly.
It follows from the definition of $\phi$ that
\begin{equation}
  h=h_{0}+\sqrt{\frac{2P}{K}\phi^{2}+(h_{B}-h_{0})^{2}},
\label{height}
\end{equation}
where $h_{B}$ is the elevation of the water table at the boundary.
A least squares fit of the measured elevations to
equation~(\ref{height}) gives estimates $P/K=7\times10^{-5}$,
$h_{0}=38$~m, and $h_{B}=38$~m (figure~\ref{GPRfig}d).
Additionally taking $P$ to be the observed mean rainfall rate of
$5\times10^{-8}$ m sec$^{-1}$, gives $K=6\times10^{-4}$ m sec$^{-1}$.
Each of these estimates is consistent with the analysis of
Ref.~\cite{abrams2009growth}.
Furthermore, the estimated permeability is consistent with the permeability of
clean sand~\cite{bear1979hydraulics}.
The elevation $h_{0}$ of the impermeable layer may be overestimated
due to uncertainties in the analysis of the GPR data.

\section{Comparison of contour curvature to the groundwater flux}

Because the curvature is a function of the second derivative of a
curve, its estimation requires an accurate characterization of the
channel shape.
To have the highest possible accuracy in the estimation of curvature,
we restrict the comparison between flux and curvature to a small piece
of the network where the boundary is linearly interpolated between
points separated by $5$ m.

\begin{figure}
  \centering
\begin{tabular}{cc}
\includegraphics[scale=.4]{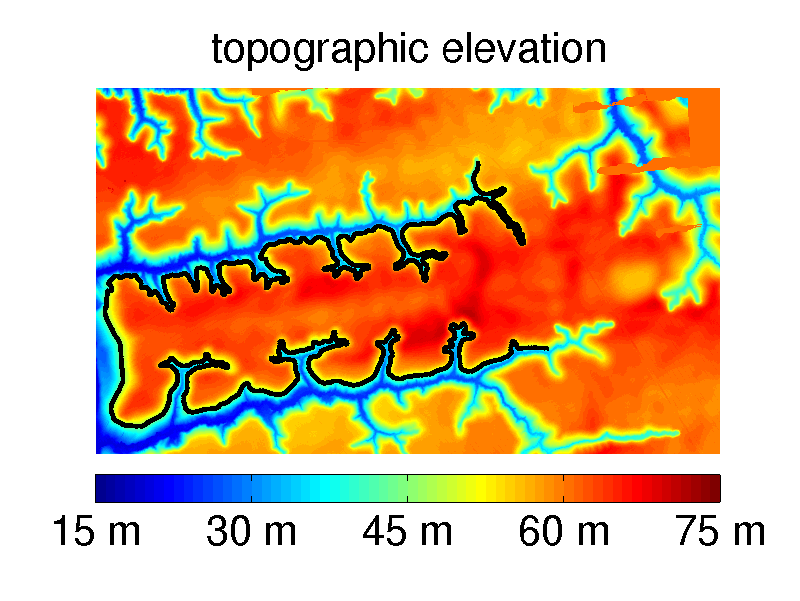}&
\includegraphics[scale=.4]{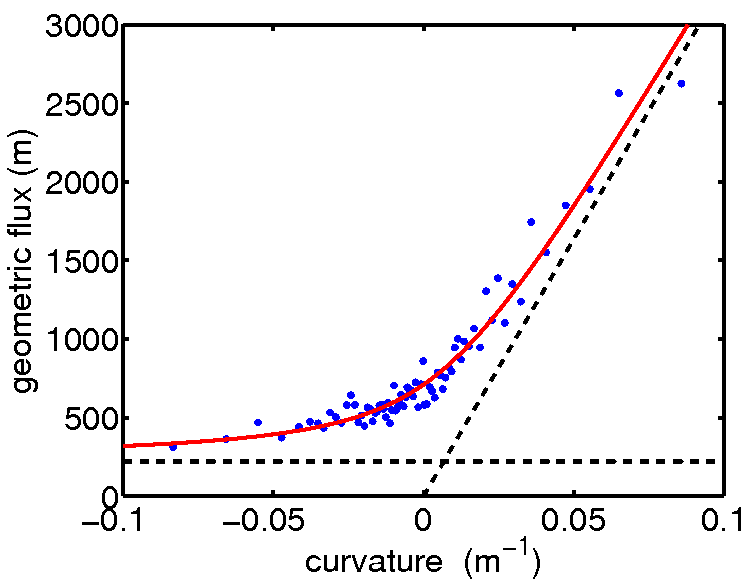}\\
(a)&(d)\\
\includegraphics[scale=.4]{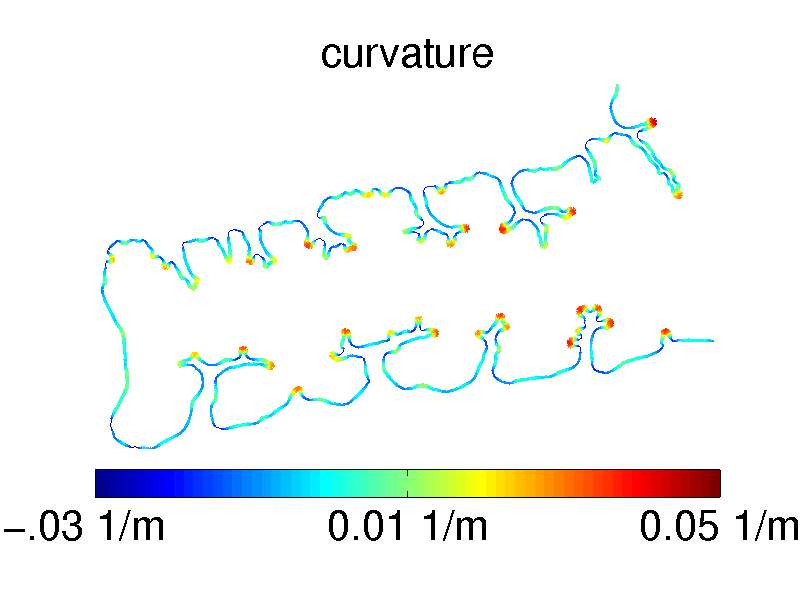}&
\includegraphics[scale=.4]{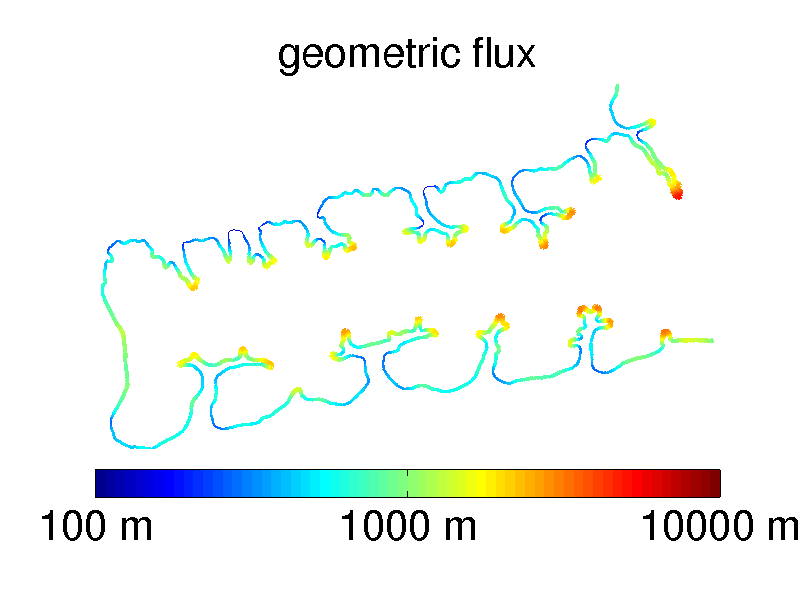}\\
(b)&(c)\\
\end{tabular}
\caption{Identification of the relationship between the curvature of
  the valley walls and the local flux of groundwater. (a) The
  curvature and flux are measured between two valleys along the black
  contour. (b) The curvature at each point on the boundary is measured
  by fitting a circle to boundary. (c) The flux into each section of
  the network is found from the solution of
  equation~(\ref{poisson_}). (d) Comparison of the flux into each
  section of the network to the curvature. Geometric reasoning gives
  the asymptotic behavior (black dashed lines) of this relation when
  the magnitude of the curvature is large.}
\label{curvaturefig}
\end{figure}

The curvature at a point on the boundary is computed by fitting a
circle to the point and its neighbors on both sides
(Figure~\ref{curvaturefig}b).
Given the best fitting circle, the magnitude of the curvature is the
inverse of the radius. The curvature is negative when the center of
the circle is outside the valley and positive when the center is inside the
valley.
%

To compare the curvature and flux at a point, we calculate the
Poisson flux $q_{p}$ into each section of this piece of the network by
solving equation~(\ref{poisson_}) between the channels
(Figure~\ref{curvaturefig}c).
We closed the boundary on the eastern side of the domain by attaching
the extremities to the valley network to the east using
zero-flux boundaries.
To identify the characteristic dependence of the flux at a point on
the curvature, we averaged the flux and curvature at points on the
boundary with similar curvatures. Each point in
Figure~\ref{curvaturefig}d represents the average flux and curvature
of 50 points on the boundary.

\section{The Poisson flux-curvature relation}

\begin{figure}[h]
\centering
\includegraphics[scale=.5]{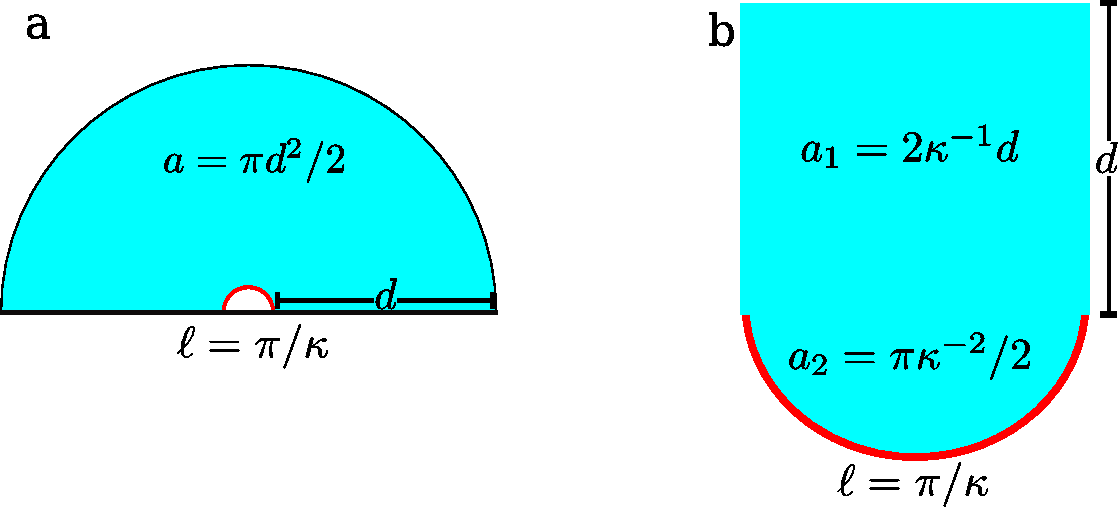}
\caption{The Poisson flux is the local drainage density. (a) When a
  basin drains into a convex region (red line) the drainage density
  increases with curvature $\kappa$. (b) When a basin drains into a
  concave interval, the drainage density decreases with curvature.}
\label{drainage}
\end{figure}

The Poisson flux is the area that drains into small segment of the
network divided by the length of the segment.
It can therefore be considered as a ``local'' inverse drainage density.
Because all of the area drains into some piece of the channel, the
integral of the Poisson flux is the total area of the basin. It
follows that its mean value is the inverse drainage density.

In what follows we ask how the Poisson flux depends on the distance
$d$ a piece of the network is from its drainage divide.
We note that if a $d$ has a characteristic value in a network, then we
find a scaling of geometric flux with curvature that is consistent
with observation (Fig. 1c).

A section of the network receives a large flux when it drains a large
area $a$ or when all of the water is forced through a small length of
channel wall $\ell$.
When water from a large basin ($d \gg \kappa^{-1}$) drains toward a
point
, then $a \sim d^{2}$ (Fig.~\ref{drainage}a).
Note that ``$\sim$'' is the symbol for ``is the order of magnitude
of'' or ``scales as.''
This area is drained into a section of channel, the length $\ell$ of
which is proportional to the planform radius of curvature,
$\kappa^{-1}$; thus in regions of high curvature
\begin{equation}
  q_{p}=\Omega\kappa,
  \label{hyperbola}
\end{equation}
where $\Omega=md^{2}$ is a constant of the network related to the
characteristic groundwater discharge of a head and $m\sim1$ is a
proportionality constant related to the characteristic shape of a
valley head.
The flux into a point is therefore proportional to the product of
variables characterizing the network, $\Omega$, and the local geometry
of the channel, $\kappa$.
Equating $d$ with the inverse drainage density of the network, we find
$d=147$ m from the analysis of the topographic map.
Fitting a hyperbola to the data in Fig.~1c, given this
value of $d$, gives $m=1.5\pm0.2$, consistent with $m\sim1$.

In concave regions of the channel the area drained is the sum of the
area outside the concavity and the area inside the concavity (Fig.~\ref{drainage}b).
This area $a$ can be expressed as
\begin{equation}
  a=m_{1}d \kappa^{-1}+m_{2}\kappa^{-2},
\end{equation}
where $m_{1}$ and $m_{2}$ are dimensionless numbers related to the
shape of the drainage basin outside and inside the concavity,
respectively.
For example, if the concavity is a semi-circular depression and it
drains a rectangular region, then $m_{1}=2$ and $m_{2}=\pi/2$.
This area is drained by a segment of length $\sim\kappa^{-1}$ giving a
mean Poisson flux $q_{p}$ that scales as
\begin{equation}
  q_{p}=(m_{1}d+m_{2}\kappa^{-1})/m_{3},
\end{equation}
where $m_{3}$ is a dimensionless number related to the shape of the
concavity.
Fitting the data to a hyperbola, and again taking $d=147$ m, we find
$m_{1}/m_{3}=1.52\pm0.22$ and $m_{2}/m_{3}=10.80\pm2.97$.
This scaling relation, in combination with the behavior at large
positive $\kappa$, gives the the asymptotic behavior of
the flux-curvature relation.

\section{Derivation of the shape of the valley head}

Here we derive equation~(5) of the main text.

The balance between translation and curvature-driven growth relates
the orientation to the curvature through the equation
\begin{equation}
\pi\cos\theta=w\kappa.
\end{equation}
We first re-write the orientation of a segment in terms of the local
normal $\hat{n}(x)$ to the curve and the direction the head is
translating $\hat{y}$. It follows from the definition of $\theta$ that
\begin{equation}
\pi\hat{n}(x)\cdot\hat{y}=w\kappa(x),
\label{curvature}
\end{equation}
Next, by describing the shape of a valley head by a curve $y(x)$,
equation~(\ref{curvature}) becomes
\begin{equation}
\frac{-\pi}{\sqrt{1+(\partial_{x}y)^{2}}}=w\frac{\partial_{xx}y}{(1+(\partial_{x}y)^{2})^{3/2}}.
\end{equation}
With the substitution $g=\partial_{x}y$, this equation is re-expressed
as an integrable, first order equation as
\begin{equation}
w\partial_{x}g+\pi(1+g^{2})=0.
\end{equation}
Integrating once,
\begin{equation}
g=\partial_{x}y=-\tan\left(\frac{\pi x}{w}\right).
\end{equation}
Integrating a second time for $y$ gives
\begin{equation}
y=\frac{w}{\pi}\log\cos\left(\frac{\pi x}{w}\right),
\label{logcos}
\end{equation}
equivalent to equation~(5) of the main text.

Although not necessary here, it is occasionally useful
to express the shape of the channel as a vector $\mathbf{v}$
parameterized by arc length $s$,
\begin{equation}
\mathbf{v}(s)=\frac{w}{\pi}\left(\begin{array}{c}
2\mbox{arctan}(\mbox{tanh}(\pi s/2 w))\\
\mbox{log}(\mbox{sech}(\pi s/w))\\
\end{array}\right).
\end{equation}
The derivative $\mathbf{v}$ is the unit tangent vector.

\section{Selection of valley heads}

The derivation of equation~(\ref{logcos}) requires that the channel grow
forward without changing shape.
Consequently, when identifying seepage valley heads suitable for
analysis, we restricted our analysis to isolated channels.

\begin{figure}[h]
\centering
\includegraphics[scale=.62]{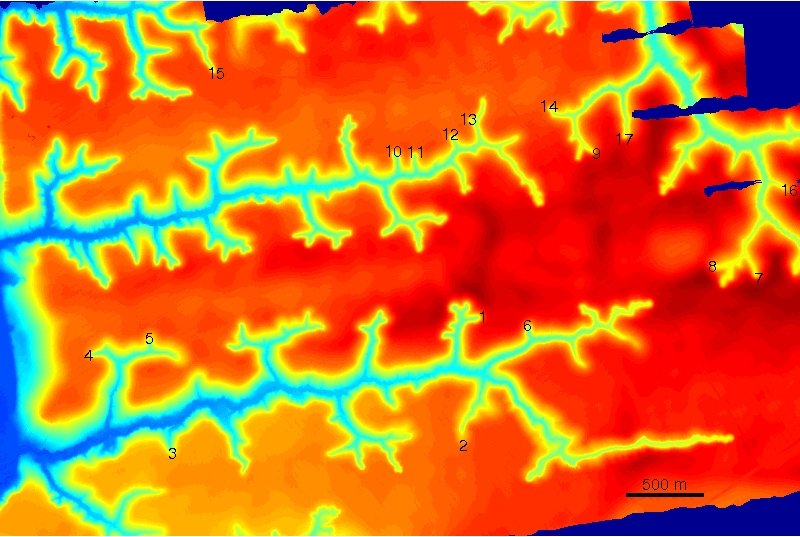}
\caption{17 isolated valley heads were chosen from the Florida
  network}
\label{map}
\end{figure}

\newpage
\subsection{Florida Network}

We select valley heads from the Florida network that are reasonably
isolated and not bifurcating. Given such a valley, we extract an
elevation contour approximately one half the distance between the
spring and the upland flat plain. We find that the deviation in the
shape of any given channel from equation~(\ref{logcos}) is insensitive
to the choice of elevation contour.


\begin{table}[h]
\centering
\caption{Valley heads from the Florida network. Coordinates are given with respect to UTM zone 16R}
\begin{tabular}{ccccc}
\centering
channel  &  Easting (m) & Northing (m) & elevation (m)  &  width  (m)  \\
\hline
\hline
1  &  696551.40  &  3373949.52  &  56.91  &  100.66 \\
2  &  696423.49  &  3373123.32  &  55.74  &  111.03 \\
3  &  694537.55  &  3373068.53  &  42.93  &  50.47 \\
4  &  693995.09  &  3373701.11  &  49.49  &  49.60 \\
5  &  694391.80  &  3373813.01  &  43.49  &  38.72 \\
6  &  696841.09  &  3373900.80  &  44.23  &  28.96 \\
7  &  698339.72  &  3374200.55  &  59.95  &  83.53 \\
8  &  698040.54  &  3374282.69  &  50.91  &  48.28 \\
9  &  697285.68  &  3375011.47  &  59.12  &  80.20 \\
10  &  695968.97  &  3375029.24  &  49.15  &  56.05 \\
11  &  696114.42  &  3375019.47  &  46.62  &  42.59 \\
12  &  696336.97  &  3375135.23  &  49.88  &  56.34 \\
13  &  696453.90  &  3375233.09  &  51.13  &  50.54 \\
14  &  696976.13  &  3375317.38  &  51.10  &  46.08 \\
15  &  694818.57  &  3375532.39  &  54.82  &  52.04 \\
16  &  698537.91  &  3374777.58  &  54.21  &  70.77 \\
17  &  697463.52  &  3375108.63  &  53.97  &  55.17 \\
\end{tabular}
\end{table}

\pagebreak[4]
\newpage
\subsection{Experiments}

The experimental apparatus used to grow seepage channels has been
previously described~\cite{schorghofer2004spontaneous}. 
The channel 
used in the comparison to equation~(\ref{logcos}) grew from an
initially rectangular indentation 3 cm deep in a bed of $0.5$ mm glass
beads sloped at an angle of $7.8^{\circ}$ with a pressure head of 19.6
cm.
To extract the shape of the channel, we first removed the slope of the
bed by subtracting the elevation of each point at the beginning of the
experiment.
We then follow the growth of an elevation contour a constant depth
below the surface.
Because the shape of channel at the beginning of the experiment is
heavily influenced by the shape of the initial indentation, we
restrict our analysis to the shape of the contour after 45 minutes of
growth. The channel grew for a total of 119 minutes and was measured
at 3 minute intervals.
%

\begin{figure}[t]
\centering
\includegraphics[scale=.7]{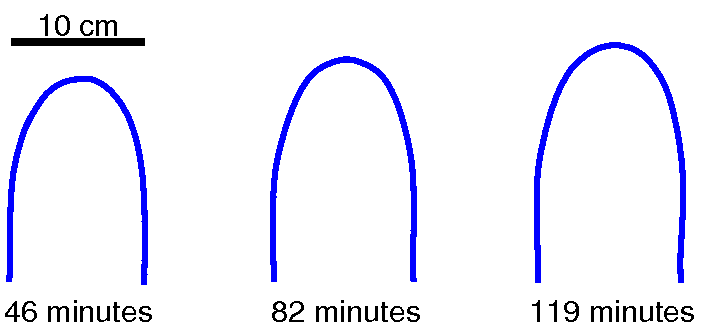}
\caption{An elevation contour (blue lines) was extracted from the
  experiment every three minutes from a digital elevation
  map~\cite{lobkovsky2007erosive}. These three representative
  elevation contours from the beginning, middle, and end of the
  experiment demonstrate that the shape changed little during growth.}
\label{experiment_ref}
\end{figure}

\pagebreak[4]
\subsection{Snake River valley heads}

\begin{figure}[h]
\centering
\includegraphics[scale=.7]{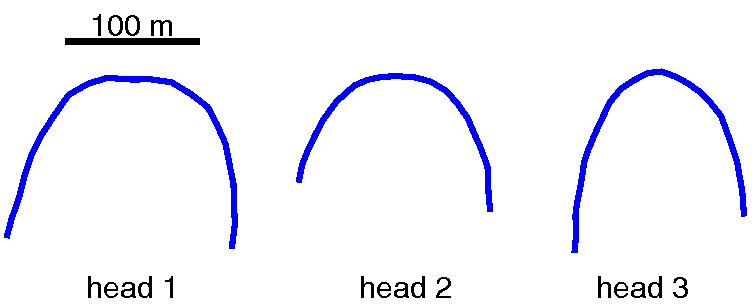}
\caption{The shape of amphitheater-shaped valley heads growing off of
  the Snake River in Idaho were extracted from aerial photos of the
  channels. Heads 1 and 2 are near Malad Gorge. Head 3 is Box
  Canyon.}
\label{snakefig}
\end{figure}

To compare the form of amphitheater-shaped valley heads growing off of
the Snake River in Idaho, we extract the valley shape from images
taken from Google Earth. We select three prominent heads
(Table~\ref{snaketab}, Figure~\ref{snakefig}); Box
Canyon~\cite{lamb2008formation} and two near Malad Gorge. We extract
the shape of each of these heads by selecting points at the upper edge
of the valley head. The mean spacing between points is 13 m. We stop
selecting points when the valley turns away from the head.

\begin{table}[h]
\centering
\caption{Valley heads near the Snake River}
\begin{tabular}{cccc}
\centering
channel & latitude & longitude  &  width  (m)  \\
\hline
\hline
1  &  42.8675$^{\circ}$  &  115.6432$^{\circ}$  &  190\\
2  &  42.8544$^{\circ}$  &  115.7045$^{\circ}$  &  166\\
3  &  42.7084$^{\circ}$  &  114.9683$^{\circ}$  &  132\\
\end{tabular}
\label{snaketab}
\end{table}

\newpage
\subsection{Martian valley heads}

\begin{figure}[h]
\centering
\includegraphics[scale=.6]{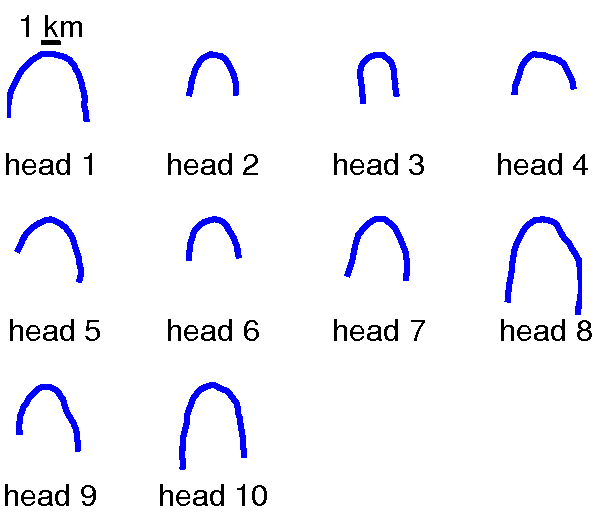}
\caption{10 valley heads near the Nirgal Valley, Mars. The shape of
  each head was extracted by selecting points at the edge of the
  valley head from images generated by the Mars Odyssey orbiter.}
\label{marsfig}
\end{figure}

The shapes of the Martian ravines which we compared to
equation~(\ref{logcos}) were extracted from images generated by the
Themis camera on the Mars Odyssey orbiter. Channels are selected based
on the condition that the amphitheater head was largely isolated from
neighboring structures. Because the ravines are deeply incised into
the topography, there is typically a sharp contrast between the
ravines and the surrounding topography. We extract the shape of the
ravine by selecting points spaced of order 100~m apart along the
edge of the ravine (Table~\ref{marstab}, Figure~\ref{marsfig}). We
stop selecting points when the ravine intersects with a neighboring
structure or when the direction of the valley curves away from the
head.

\begin{table}[h]
\centering
\caption{Martian valley heads}
\begin{tabular}{ccccc}
\centering
head & Themis Image &  latitude & longitude  &  width  (m)  \\
\hline
\hline
1  &  V06395001   &   -8.7270$^{\circ}$  &  278.1572$^{\circ}$ & 4730\\
2  &  V06395001   &   -8.7235$^{\circ}$  &  278.1557$^{\circ}$ & 2650\\
3  &  V09004001   &   -9.4310$^{\circ}$  &  274.6110$^{\circ}$ & 1940\\
4  &  V11138002   &   -7.9183$^{\circ}$  &  275.4740$^{\circ}$ & 3690\\
5  &  V11138002   &   -7.9160$^{\circ}$  &  275.4736$^{\circ}$ & 3740\\
6  &  V14133002   &   -9.5763$^{\circ}$  &  278.4435$^{\circ}$ & 2940\\
7  &  V14857001   &   -7.5656$^{\circ}$  &  273.6060$^{\circ}$ & 3110\\
8  &  V16654002   &   -8.7792$^{\circ}$  &  275.5868$^{\circ}$ & 3970\\
9  &  V16654002   &   -8.7781$^{\circ}$  &  275.5894$^{\circ}$ & 3310\\
10 &  V26750003   &   -8.0633$^{\circ}$  &  274.8977$^{\circ}$ & 3370\\
\end{tabular}
\label{marstab}
\end{table}

\newpage
\section{Stream discharge data }

\subsection{Comparison of field measurements to the predicted flux}

Fig.~1 of the main text compares the solution of
equation~(\ref{poisson_}) to field measurements. The instantaneous
discharge of a stream is measured from the the cross-sectional area
$a$ in a locally straight section of the channel and the surface
velocity $v$, from which the discharge $Q=av$. We measure the surface
velocity of the stream from the travel time of a small passive tracer
between points at a fixed distance. This method may underestimate the
discharge in very small streams where a substantial fraction of the
flow may be moving through the muddy banks of the stream.

To compare the measured discharge to the Poisson equation, we
integrate the flux, $q=P\|\nabla\phi^{2}\|$, along the section of the
network upstream from the measurement assuming the reported annual
rainfall, $P=5\times10^{-8}$ m sec$^{-1}$. When discharge is measured
near a spring, the flux is integrated around the valley head.

\subsection{January 2009}
\begin{center}
\begin{tabular}{cccc}
  Easting (m) & Northing (m)  & discharge (cm$^{3}$ sec$^{-1}$) & predicted discharge (cm$^{3}$ sec$^{-1}$)\\
  \hline
  \hline
  696905.45  &   3374708.15   &   11700   &   4802.54\\
  695425.47  &   3374595.35   &   310   &   422.30\\
  695333.80  &   3374486.16   &   1900   &   1459.50\\
  695410.03  &   3374422.51   &   2100   &   1403.02\\
  695589.53  &   3374413.20   &   2000   &   1272.90\\
  695608.45  &   3374439.65   &   1700   &   1145.17\\
  695602.26  &   3374467.29   &   4700   &   2103.67\\
  695532.14  &   3374764.77   &   310   &   490.24\\
  694045.68  &   3373713.71   &   710   &   1708.72\\
  694102.24  &   3373742.47   &   850   &   1381.31\\
  694110.98  &   3373726.59   &   850   &   3520.83\\
  694393.38  &   3373788.44   &   810   &   1761.83\\
  694515.20  &   3373714.30   &   2300   &   2051.40\\
%
   694700.99  &   3373494.69   &   2900   &   1831.06\\
   697174.63  &   3373662.18   &   700   &   3004.36\\
   697622.18  &   3374045.11   &   10800   &   4700.85\\
   697523.57  &   3374034.52   &   440   &   2225.53\\
   696432.08  &   3373937.74   &   3500   &   2619.48\\
   696353.61  &   3374006.59   &   3500   &   2688.01\\
   696415.16  &   3373979.53   &   3600   &   2545.77\\
   696363.79  &   3373884.98   &   570   &   673.20\\
   696314.56  &   3373838.46   &   3100   &   1132.34\\
   695400.74  &   3373894.43   &   2800   &   2117.42\\
   695417.69  &   3373884.87   &   3100   &   2522.08\\
   694429.25  &   3374329.77   &   700   &   1145.16\\
   694541.01  &   3374318.70   &   1250   &   1626.34\\
   694295.68  &   3374320.27   &   700   &   950.41\\
   694081.94  &   3374205.31   &   1950   &   2969.80\\
   693696.69  &   3373094.27   &   300000   &   284251.54\\
   693575.95  &   3374496.41   &   100000   &   148834.48\\
\end{tabular}
\end{center}

\subsection{April 2009}

\begin{center}
\begin{tabular}{cccc}
 Easting (m) & Northing (m)  & discharge (cm$^{3}$ sec$^{-1}$) & predicted discharge (cm$^{3}$ sec$^{-1}$)\\
\hline
\hline
   696577.00  &   3375064.00   &   26688   &   17558.21\\
   696515.00  &   3375060.00   &   31324   &   25683.40\\
   696526.00  &   3375081.00   &   10878   &   7960.94\\
   696378.00  &   3375075.00   &   3155   &   2196.50\\
   696374.00  &   3375047.00   &   31168   &   28808.33\\
   696312.00  &   3374949.00   &   44714   &   33750.34\\
   693684.21  &   3374490.78   &   181142   &   148233.29\\
   693857.70  &   3374480.17   &   134261   &   134367.19\\
   694237.23  &   3374550.07   &   159510   &   123111.22\\
   694371.93  &   3374575.91   &   96597   &   120688.53\\
   694445.52  &   3374574.44   &   123230   &   115351.89\\
   694706.00  &   3374606.66   &   142841   &   111689.91\\
   694808.19  &   3374666.05   &   133061   &   103032.24\\
   694815.26  &   3374674.12   &   24251   &   14558.59\\
   695449.36  &   3374792.20   &   70782   &   70543.57\\
   695317.14  &   3374776.63   &   115771   &   82714.80\\
   695400.81  &   3374783.02   &   18354   &   10476.11\\
   695613.16  &   3374808.59   &   46630   &   69195.87\\
   695756.20  &   3374863.59   &   11422   &   10024.07\\
   695787.02  &   3374851.22   &   81630   &   57339.57\\
   695914.95  &   3374827.10   &   24757   &   41590.16\\
   695922.74  &   3374822.92   &   31480   &   15071.30\\
   696011.72  &   3374871.04   &   6903   &   2472.52\\
   696019.12  &   3374873.43   &   52090   &   38588.88\\
   696127.23  &   3374876.96   &   44644   &   36223.61\\
   696267.06  &   3374905.73   &   51745   &   34747.90\\
   696335.93  &   3374970.34   &   52171   &   29296.08\\
%
   696577.00  &   3375064.00   &   26688   &   17558.21\\
   696515.00  &   3375060.00   &   31324   &   25683.40\\
   696526.00  &   3375081.00   &   10878   &   7960.94\\
   696378.00  &   3375075.00   &   3155   &   2196.50\\
   696374.00  &   3375047.00   &   31168   &   28808.33\\
   696346.42  &   3374960.48   &   8141   &   4287.83\\
   696916.41  &   3374703.37   &   2704   &   2773.16\\
   696913.37  &   3374697.05   &   2131   &   1230.01\\
   695406.38  &   3373894.53   &   6791   &   2117.42\\
   695284.53  &   3373820.41   &   4975   &   6209.90\\
   695268.73  &   3373828.01   &   12171   &   7245.29\\
   695207.06  &   3373539.81   &   28499   &   16435.07\\
   695163.07  &   3373472.73   &   285299   &   214562.12\\
   695825.64  &   3373844.08   &   20009   &   6515.55\\
\end{tabular}
\end{center}

\begin{center}
\begin{tabular}{cccc}
  Easting (m) & Northing (m)  & discharge (cm$^{3}$ sec$^{-1}$) & predicted discharge (cm$^{3}$ sec$^{-1}$)\\
  \hline
  \hline
  695818.15  &   3373874.10   &   3098   &   1747.84\\
  695829.40  &   3373872.17   &   6847   &   4724.56\\
  695870.78  &   3373925.80   &   1292   &   1436.14\\
  695873.99  &   3373937.31   &   7298   &   2562.84\\
  694804.18  &   3374918.55   &   4777   &   2926.35\\
  694811.43  &   3374929.34   &   15554   &   10799.79\\
  694864.00  &   3374985.35   &   9906   &   7942.65\\
  694853.04  &   3375015.85   &   6866   &   2769.17\\
  694999.62  &   3375057.93   &   11789   &   6421.11\\
  695043.10  &   3375092.62   &   3376   &   2175.13\\
  695043.00  &   3375070.20   &   5248   &   3776.13\\
  695410.00  &   3373885.00   &   4173   &   2590.45\\
  697528.00  &   3374024.00   &   995   &   2400.35\\
  695529.00  &   3374749.00   &   685   &   490.24\\
  695437.00  &   3374602.00   &   263   &   422.30\\
  695434.00  &   3374600.00   &   10759   &   9777.40\\

\end{tabular}
\end{center}
\bibliography{chan_bib.bib}{}